\count100=423
\magnification=\magstep1
 \baselineskip=12.67pt plus 2pt minus 1 pt 
 \hsize=6truein
 \hoffset=18pt
 \parskip=1pt plus 1pt
 \vsize=8truein
 \def\nullx{\hfill}
 \pageno=\count100
 \headline{\ifodd\pageno\rightheadline \else\leftheadline\fi}
 \footline{\ifnum\pageno=\count100{\hss}\else\nullx\fi}
 \def\rightheadline{\ifnum\pageno=\count100 \nullx%
\else\it\chptitle\hfil\rm\folio\fi}
 \def\leftheadline{\ifnum\pageno=\count100 \nullx%
\else\rm\folio\hfil\it\author\fi}
 \hfuzz=1pt
 \tolerance=10000

 
 \font\title=cmbx12

 \font\sc=cmcsc10

 \newif\ifninebanner\ninebannertrue 
 \newdimen\bannerskip
 \ifninebanner
 \bannerskip=8.5pt
 \font\brm=cmr9 scaled 833
 \font\bbf=cmbx9 scaled 833
 \font\bsl=cmsl9 scaled 833
\font\bsy=cmsy9 scaled 833
 \else  
 \bannerskip=7.5pt
 \font\brm=cmr8 scaled 833
 \font\bbf=cmbx8 scaled 833
 \font\bsl=cmsl8 scaled 833
\font\bsy=cmsy8 scaled 833
 \fi

\def\eth{{\mathsurround=0pt $\delta$\kern-0.6em\raise0.6ex\hbox{$-$}}}
\def\thorn{{\mathsurround=0pt $\flat$\kern-0.44em\lower0.17ex\hbox{$\vert\;$}}}

\def\beginflushfootnote{\bgroup\parindent=0pt\footnote}
 \def\endflushfootnote{\egroup}

\hfuzz=1pt
\def\n{\noindent}
\def\chptitle{Symmetries  of Heterotic String Effective Theory}
\def\author{D.V.\ Gal'tsov}
\def\frac#1#2{{#1 \over #2}}

\null\vskip72pt

\centerline{\title Symmetries of Heterotic String Effective }\smallskip
\centerline{\title Theory in Three
and Two Dimensions$^1$ }\vskip12pt

\centerline{\sc D.V. Gal'tsov}\smallskip
\centerline{\sl  Department of Theoretical Physics}
\centerline{\sl Moscow State University}
\centerline{\sl Moscow 119899}
\centerline{\sl Russia}\vskip24pt


{\narrower \bigskip\noindent {\bf Abstract.}
The four-dimensional bosonic effective action of the toroidally compactified
heterotic string incorporating a dilaton, an axion and one $U(1)$ vector
field is studied on curved space-time manifolds with one and two commuting 
Killing vectors. In the first case the theory is reduced to a 
three-dimensional sigma model possessing a symmetric pseudoriemannian 
target space isomorphic to the coset $SO(2,3)/(SO(3)\times SO(2))$. 
The ten-parameter group $SO(2,3)$ of target space isometries contains 
embedded both $S$ and $T$ classical duality symmetries of the heterotic 
string. With one more ignorable coordinate, the theory reduces to a 
two-dimensional chiral model built on the above coset, and therefore 
belongs to the class of completely integrable systems. This entails 
infinite-dimensional symmetries of the Geroch--Kinnersley--Chitre type. 
Purely dilatonic theory is shown to be two-dimensionally
integrable only for two particular values of the dilaton coupling constant.
In the static case (diagonal metrics) both theories essentially coincide;
in this case the integrability property holds for all values of the dilaton
coupling.\bigskip}

\beginflushfootnote{}{{\baselineskip\bannerskip\brm
{\bsl \noindent $^1$ A talk given at the International Workshop 
``Heat Kernel Techniques and Quantum Gravity'', Winnipeg, 
Canada, 2---6 August, 1994), published in
{\bbf Heat Kernel Techniques and Quantum Gravity},
ed.\ by  S.~A. Fulling,
 {Discourses in Mathematics and Its Applications},
 No.~4, \textfont2=\bsy \copyright\
 Department of Mathematics, Texas A\&M University,
 College Station, Texas, 1995, pp.~423--449.
\par}}\endflushfootnote

\vfill\eject

\n {\bf 1. Introduction}

A well-known property of the Einstein equations in General Relativity,
crucial for the opportunity to find many nontrivial
exact solutions to this theory, is two-dimensional integrability,
which is manifest when the theory is restricted to space-times
possessing two commuting Killing vector fields [1].
Because of this property, in principle all solutions with two ignorable
coordinates can be found using various solution-generating methods.
Although the problem of constructing  solutions having desired physical
properties remains rather complicated technically, with modern
mathematical tools it is becoming more and more tractable. As a matter
of fact, the existing knowledge of sufficiently general classes of
solutions is essentially related to this property;
other methods to solve highly nonlinear Einstein equations are either
restricted to more symmetric cases, or able to produce only some
particular types of solutions [2].

It has been observed that  a similar integrability property
is shared by other theories
related to General Relativity: Einstein--Maxwell (EM) theory
[3], higher dimensional vacuum Einstein equations [4],
bosonic sectors of some dimensionally reduced supergravity theories
[5], [6], [7]. The crucial feature ensuring two-dimensional
integrability is the symmetric-space property of the corresponding
sigma model in three dimensions (for a review see, e.g.,
Breitenlohner and Maison [6]).
Both Einstein and massless vector field equations, being reduced to three
dimensions via an imposition of a non-null space-time Killing symmetry,
may be described in terms of gravity coupled sigma models. If additional
scalar fields present do not destroy this property, the
main question is whether the target space is a space of a constant
Riemann curvature. If yes, the equations of motion may be cast into the
form of modified chiral matrix equations. The corresponding Lax pair
is readily given when the theory is further reduced to two dimensions.
The symmetric-space property in its turn requires that the target space
possess a sufficient number of isometries. This convenient way
to investigate the integrability property is basically due to the
idea of potential space introduced long ago by Neugebauer and Kramer
[8].

It seems likely that the description of gravity at  ultramicroscopic
distances is in the scope of  string theory, the most promising
being the heterotic string model. In the low-energy (classical)
limit of this theory with compactified extra dimensions one gets
in the bosonic sector
an Einstein gravity coupled to massless vector and scalar fields.
If initially formulated on a $D$-dimensional curved
space-time manifold possessing $d$ Abelian isometries with $p$
initial $U(1)$ vector fields present, such a theory is known  to
possess global invariance under an $SO(d,d+p)$ group called
$T$-duality [9], [10]. In four dimensions there is another
important symmetry, $S$-duality [11], whose occurrence is related
to the possibility of representing a Kalb--Ramond field by a scalar
Peccei--Quinn axion. Classical $S$-duality is the $SL(2,R)$ group
containing electric-magnetic continuous rotations as well as
the dilaton transformation ensuring a connection between weak and
strong coupling regimes of the string theory.
It is conjectured that discrete subgroups
of these dualities are exact symmetries of the heterotic string.

The simplest model of this kind, often called dilaton-axion gravity,
which incorporates basic features of the full effective action, can be
formulated directly in $D=4$ where it includes one $U(1)$ vector ($p=1$)
and two scalar fields coupled in a way prescribed by the
heterotic string effective theory:
$$S=\frac{1}{16\pi}\int \left\{-R+2\partial_\mu\phi\partial^\mu\phi +
\frac{1}{2} e^{4\phi}
{\partial_\mu}\kappa\partial^\mu\kappa
-e^{-2\phi}F_{\mu\nu}F^{\mu\nu}-\kappa F_{\mu\nu}{\tilde F}^{\mu\nu}\right\}
\sqrt{-g}\, d^4x.\eqno (1.1)$$
Here $F=dA,\;{\tilde F}^{\mu\nu}=
\frac{1}{2}E^{\mu\nu\lambda\tau}F_{\lambda\tau}\,$,
and $\phi,\; \kappa$ are dilaton and axion fields.
This model may be regarded as the Einstein--Maxwell theory
coupled to dilaton and axion (EMDA theory).
The purpose of my talk is to show that this theory possesses
the same integrability property as  pure EM theory, while the underlying
algebraic structure is rather different. The investigation has two goals.
Firstly, the study of further dimensional reduction to three and two
dimensions may reveal new hidden symmetries of the heterotic string. In fact,
for $d=p=1$, $T$-duality is $SO(1,2)$, and together with $SL(2,R)$
$S$-duality this is insufficient to ensure the desired symmetric-space
property of the target space. Hence, the question of existence of such
a property is closely related to the possibility of
embedding of $T$ and $S$ dualities into a larger group.
Secondly, if it exists, the integrability property could be very helpful in
constructing exact classical solutions
to the action (1.1), which is important both for extracting gravitational
predictions of the string theory, and for a search of new quantum
four-dimensional string models on $D=4$ curved space-time manifolds.
Some of the results reported here were
obtained in collaboration with O.V.~Kechkin [12].\bigskip

\n{\bf 2. Potential space }

Both three-dimensional (finite) and
two-dimensional (infinite) symmetries are closely related for the
class of two-dimensionally integrable systems, so we  start with
reducing (1.1) to three dimensions, imposing a non-null
Killing symmetry on the $D=4$ manifold. In the case of a
timelike Killing vector, the
four-dimensional metric can be presented
in terms of a three-metric $h_{ij}$, a rotation one-form
$\omega_i\; (i, j=1, 2, 3)$,
and a three-dimensional conformal factor $f$ depending only
on the 3-space coordinates $x^i$:
$$ds^2=g_{\mu\nu}dx^\mu
dx^\nu=f(dt-\omega_idx^i)^2-\frac{1}{f}h_{ij}dx^idx^j. \eqno (2.1)$$

The subsequent derivation of the three-dimensional $\sigma$ model
is rather standard and mainly follows Israel and Wilson [13].
A four-dimensional vector field in three dimensions is represented by two
scalars. In our case of timelike Killing symmetry they have the meaning of
electric $(v)$ and magnetic $(u)$ potentials and can be introduced via
relations $$F_{i0}=\frac{1}{\sqrt{2}}\partial_iv,\eqno (2.2)$$
$$e^{-2\phi}F^{ij}+\kappa {\tilde F}^{ij}=\frac{f}{\sqrt{2h}}\epsilon^{ijk}
\partial_ku,\eqno (2.3)$$
solving the  spatial part of the  modified Maxwell equations and
Bianchi identity. The rotation one-form $\omega_i$ in (2.1)
also reduces to a scalar --- the twist potential $\chi$:
$$\tau^i=-f^2\frac{\epsilon^{ijk}}{\sqrt{h}}\partial_j\omega_k,\;\;
\tau_i=\partial_i\chi +v\partial_iu-u\partial_iv.\eqno (2.4)$$

Therefore in three dimensions we have three pairs of scalar variables:
one $(f,\;\chi)$ is inherited from the four-dimensional metric,
the second $(v,\;u)$  from the vector field, and the last is the
dilaton-axion pair $(\phi, \kappa)$.
A typical feature of the four-dimensional gravitationally coupled
system of $U(1)$ vector fields (possibly interacting with scalars
representing some coset) is that in three
dimensions they form a gravity coupled sigma model; well-known examples
are provided by the Einstein--Maxwell system [8]
and multidimensional
vacuum Einstein equations compactified to four dimensions [4].
In our case it can be  checked directly that the equations for
$f, \chi, v, u, \phi, \kappa$ may be
obtained by variation of the following three-dimensional action:
$$\eqalignno{
S &= \int\Bigl\{{\cal R}-\frac{1}{2f^2}[(\nabla f)^2+(\nabla\chi +
v\nabla u-u\nabla v)^2]-2(\nabla\Phi )^2\cr
&\quad -\frac{1}{2}e^{4\phi}(\nabla\kappa )^2+\frac{1}{f}\left[e^{2\phi}
(\nabla u-\kappa\nabla v)^2+
e^{-2\phi}(\nabla v)^2\right]\Bigr\}\sqrt{h}\,d^3x,&(2.5)}$$
where ${\cal R}\equiv {\cal R}_i^i$, ${\cal R}_{ij}$ is the
three-dimensional Ricci tensor, and $\bf\nabla$ stands for the
3-dimensional covariant derivative.
This action  can be rewritten as the gravity-coupled three-dimensional
$\sigma$ model
$$S=\int \left({\cal R}-{\cal G}_{AB}\partial_i\varphi^A\partial_j\varphi^B
h^{ij}\right)\sqrt{h}\,d^3x,\eqno (2.6)$$
where $\varphi^A=(f,\; \chi,\; v,\;u,\;\kappa,\; \phi),\; A=1,\ldots, 6$.
The corresponding target space metric reads
$$\eqalignno{{dl}^2 & =  {\cal G}_{AB}d\varphi^A d\varphi^B =
\frac{1}{2}f^{-2}[{df}^2+(d\chi +vdu-udv)^2]\cr
&\quad -f^{-1}[e^{2\phi }{(du-\kappa dv)}^2  +e^{-2\phi }{dv}^2] +
2{d\phi }^2+\frac{1}{2}e^{4\phi }{d\kappa }^2.&(2.7)}$$
This is a six-dimensional pseudoriemannian space with indefinite metric,
which generalizes Neugebauer and Kramer's potential space
found for the stationary Einstein--Maxwell system in 1969 [8].
It is worth noting that the present $\sigma$ model does not reduce to the
Einstein--Maxwell one when $\kappa=\phi=0$, because the equations for
a dilaton  and an axion  generate constraints $F^2=F\tilde F =0$. Hence,
generally the solutions
of the Einstein--Maxwell theory with one Killing symmetry are not related
to solutions of the present theory by target-space isometries (except for
the case $F^2=F\tilde F =0$). On the other hand, one can consistently set
in the present $\sigma$ model $v=u=\phi=\kappa=0$, reducing it to the
Einstein vacuum sigma model. Therefore all solutions to the vacuum
Einstein equations with one non-null Killing symmetry
are related to some solutions of the system in question by target space
isometries. This fact was used in [12] to generate the most general
family of black hole solutions to dilaton-axion gravity.

In connection with other symmetric gravity-coupled systems another
model is worth  discussing:  Einstein--Maxwell--dilaton (EMD) theory
described by the action
$$S = \frac{1}{16\pi} \int \left\{-R +
2(\partial \phi)^2 -  e^{-2\alpha \phi} F^2\right\} \sqrt{-g}\, d^{4}x.
\eqno (2.8)$$
It can be regarded as a truncated version of  (2.1) without an axion
field (although more general in the sense of arbitrariness of the dilaton
coupling constant  $\alpha$).
The model is interesting both in the context of  string theory,
and also as an interpolation between two theories enjoying a
two-dimensional integrability property. For $\alpha=0$ the
action (2.8) describes the EM system coupled to a scalar field, which
is equivalent to the Brans--Dicke--Maxwell (BDM) theory with the
Brans--Dicke constant $\omega=-1$. For $\alpha=\sqrt{3}$
it corresponds to the five-dimensional Kaluza--Klein (KK) theory
in the Einstein frame. In the context of  string theory another
particular value of $\alpha$ is relevant: $\alpha=1$. It was observed
recently by Horowitz [14] that in this case some nontrivial
transformations of the Harrison type [15] can be  applied
to the Schwarzschild metric to generate  dilaton black
holes [16] (see also [17]).

Repeating the above reasoning one finds the following
potential-space metric (now five-dimensional):
$$ dl^2 = \frac{df^2
+(d\chi+vdu-udv)^2}{2 f^2} -\frac{e^{-2 \alpha \phi} dv^2 +
e^{2 \alpha \phi} du^2}{f} + 2d\phi^2.\eqno (2.9)$$
Essentially the same potential space was studied long
ago by Neugebauer [18].

An important distinction between these two target spaces is the
nondiagonal structure of the electromagnetic sector in (2.7) in contract
to the diagonal one in (2.9). Physically this is due to the mixing
of electric and magnetic components of the vector field
by an axion; obviously the electric-magnetic
duality associated with (2.7) should be non-Abelian. Similar duality
for (2.9) turns out to hold only if the dilaton is decoupled
(i.e., for $\alpha=0$), and it is Abelian in that case. Electric-magnetic
duality for (2.7) is just the notable $S$-duality $SL(2,R)$ [11] of
the string theory. The question naturally arises, which isometries of
(2.7) correspond to another symmetry of the toroidally compactified
heterotic string --- $T$-duality [10], [9], which in the
present context ($p=d=1$) should be $SO(1,2)$ (i.e.,
have the same structure as $S$-duality). As we have
shown recently [12], this symmetry is intimately related to
the Ehlers--Harrison-type transformations associated with the
action (1.1), which generalize in some nontrivial way the original
Ehlers [19] and Harrison [15] transformations known in General
Relativity.

As we will see [12], both $T$ and $S$
dualities are actually embedded into a larger ten-parameter isometry
group of (2.7). This indicates that string duality symmetries are likely
to be enhanced when the theory is reduced to three dimensions. Moreover,
this embedding ensures that the target space (2.7) is a symmetric
Riemannian space, and consequently the symmetries will be infinitely
enhanced when the theory is further reduced to two dimensions [25].
Contrary to this, it turns out that the EMD target space (2.9) generally
does not possess enough isometries to ensure such properties, unless
$\alpha=0$ (which is the BDM case) or $\alpha=\sqrt{3}$ (KK case);
in both these  critical cases one has two-dimensional integrability.
Thus the role of the axion
field in the action (2.1) (which corresponds to $\alpha=1$) is very
nontrivial in generating more hidden symmetries.

Between EMDA and EMD models there is another interesting link. If the
space-time (2.2) is {\it static}, i.e., $\omega_i=0$, then it is legitimate
to consider in (2.9) purely electric ($u=0$) and purely magnetic
($v=0$) subspaces independently. At the same time, in this case it
is consistent to set the axion field in (2.7) to zero provided we deal
with either purely electric or purely magnetic fields. Then static
truncation of the EMD action will be more general (because $\alpha$
is arbitrary). Meanwhile, since static no-axion truncation of (2.7)
is inherited from a symmetric space and it corresponds
to the noncritical $\alpha$ in (2.9), this indicates that in the static
case the EMD theory should be two-dimensionally integrable for (at least
one) noncritical dilaton coupling constant value  (as we will see,
indeed for all $\alpha$). This explains the occurrence of the Harrison
transformations in the diagonal dilaton gravity [14].\bigskip

\n {\bf 3. Unification of $T$ and $S$ dualities in three dimensions}

Once the metric of the potential space is found, the isometries
can be explored by solving the Killing equations
$$K_{A;B}+K_{B;A}=0,\eqno (3.1)$$
where covariant derivatives refer to the target space metric (2.7).
It turns out that there are ten independent Killing vectors fields
for this metric. We list  them in the order of increasing complexity.
The simplest symmetry is just $\chi$--translation, $\chi=\chi_0+g,\; g$
being a real parameter, which physically
is pure (gravitational) gauge (metric remains unchanged):
$$ {K}_g=\partial_{\chi }\,.\eqno (3.2)$$
Similarly, there are two translations of electric and magnetic potentials,
accompanied by suitable transformations of the axion:
$$v=v_0+e, \quad \chi =\chi_0-u_0 e,\eqno (3.3)$$
$$u=u_0+m,\quad\chi =\chi_0+v_0 m,\eqno (3.4)$$
(with $e$ and $m$ real parameters), the corresponding generators being
$${K}_e=\partial_v-u\partial_{\chi }\,,\eqno (3.5)$$
$$ {K}_m=\partial_u+v\partial_{\chi }\,.\eqno (3.6)$$
They are also pure gauge (electric and magnetic). Also rather simple
is the scale transformation which can be expected from  power counting
in (2.7):
$$f=f_0e^{2s}, \quad \chi =\chi_0e^{2s},\quad
v=v_0 e^{s}, \quad u=u_0e^{s}.\eqno (3.7)$$
It leaves $\kappa $ and  $\phi$ unchanged so that
$$ {K}_s=2f\partial_f+2\chi \partial_{\chi }+v\partial_v+u\partial_u\,.
\eqno (3.8)$$

 Three Killing vectors correspond to the dilaton-axion
$S$-duality subalgebra, including two rotations,
$$ {K}_{d_1}=\partial_{\kappa }+v\partial_u\,,\eqno (3.9)$$
$$ {K}_{d_2}=(e^{-4\phi }-{\kappa }^2)\partial_{\kappa }+
\kappa \partial_ \phi +u\partial_v\,,\eqno (3.10)$$
and a dilaton shift accompanied by suitable rescaling of other
variables,
$$ {K}_{d_3}=\partial_\phi -2\kappa \partial_{\kappa
}+v\partial_v-u\partial_u\,. \eqno (3.11)$$
These three satisfy $sl(2,R)$ commutation relations
$$\left[K_{d_3},K_{d_1}\right]=2K_{d_1},\quad
\left[K_{d_3},K_{d_2}\right]=-2K_{d_2},\quad
\left[K_{d_1},K_{d_2}\right]=K_{d_3}.\eqno (3.12)$$
The corresponding finite transformations read
$$v=v_0,\quad u=u_0+v_0 d_1, \quad
z=z_0+d_1,\eqno (3.13)$$
$$u=u_0, \quad v=v_0+u_0 d_2,\quad z^{-1}=z_0^{-1}+d_2,\eqno (3.14)$$
$$z=e^{-2d_3} z_0,\quad
u=u_0e^{-d_3}, \quad v=v_0e^{d_3},\eqno (3.15)$$
where $z=\kappa +ie^{-2\phi}$ is the complex dilaton-axion field
($s, d_1, d_2, d_3$ are real parameters).

A nontrivial part of the isometry group includes
a conjugate pair of ``charging'' transformations
(identified in [12]  as  Harrison-type transformations
by exploring their action
on solutions of the vacuum Einstein's equations):
$$\eqalignno{{K}_{H_1} &=  2vf\partial_f+v\partial_\phi+2w\partial_{\kappa
}+ (v^2+fe^{2\phi })\partial_v \cr
&\quad +  (\chi+uv+\kappa fe^{2\phi })\partial_u+
(v\chi +wfe^{2\phi })\partial_{\chi }\,,&(3.16)\cr
{K}_{H_2} &= 2uf\partial_f+(\kappa v-w)\partial_\phi+2(\kappa
w + ve^{-4\phi })\partial_{\kappa }+
(uv-\chi +\kappa fe^{2\phi })\partial_v \cr
&\quad +  (u^2+fe^{-2\phi }+{\kappa }^2fe^{2\phi })\partial_u+
(u\chi -vfe^{-2\phi }+\kappa wfe^{2\phi })\partial_{\chi }\,,&(3.17)}$$
where $w=u-\kappa v$.
The corresponding finite transformations are more tricky.
The first
(electric) leaves invariant the  quantities
$$f e^{-2\phi}\equiv f_0 e^{-2\phi_0},\quad  \tilde \chi=\chi -
wv\equiv \chi_0 - w_0 v_0,\eqno (3.18)$$
while the other variables transform as
$$ w=w_0+\tilde\chi_0h_1, \quad
\kappa =\kappa_0+2w_0 h_1+\tilde\chi_0 h_1^2, $$
$$(\sqrt{f}e^{\phi}\pm v)^{-1}=(\sqrt{f_0}e^{\phi_0}\pm v_0)^{-1}\mp h_1.
\eqno (3.19)$$
The second (magnetic) also leaves two combinations invariant:
$$ q= f^{-1/2} |z| e^\phi\equiv f_0^{-1/2} |z_0| e^{\phi_0},\quad
 p= f^{-1}u^+ u^- \equiv f_0^{-1}u_0^+ u_0^- ,\eqno (3.20)$$
where
$$ u^{\pm}=u\pm qf=\frac{u^{\pm}_0}{1-h_2u^{\pm}_0}\, ,\eqno (3.21)$$
while the other transformations  read
$$\eqalignno{\chi &=k_+u^+ + k_- u^-+kqf,\quad v=k_+ \frac{u^+}{u^-} + k_-
\frac{u^-}{u^+}\,,\cr
z &= \frac{fq^2}{dq-i}\,,\quad d=k+k_+ \frac{u^+}{u^-} - k_-
\frac{u^-}{u^+}\,, &(3.22)\cr
 k_{\pm}&=\frac{u^{\mp}_0}{2u^{\pm}_0}\left(v_0\pm\frac{\kappa_0f_0e^{2\phi
_0} - \chi_1}{2qf_0}\right), \quad k=\frac{\kappa_0f_0e^{2\phi_0} +
\chi_1}{2qf_0}\,, }$$
where $\chi_1=\chi_0-u_0v_0$.

The last, Ehlers-type [19] generator, which closes
the full isometry algebra,
$$\left[K_{H_1},K_{H_2}\right]=2K_{E},\eqno (3.23)$$
reads
$$\eqalignno{{K}_{E} &= 2f\chi
\partial_f+wv\partial_\phi+(w^2-v^2e^{-4\phi
})\partial_{\kappa }+ (v\chi +wfe^{2\phi })\partial_v &(3.24)\cr
&\quad +  (u\chi -vfe^{-2\phi }+\kappa wfe^{2\phi })\partial_u+
({\chi }^2-f^2+fv^2e^{-2\phi } +fw^2e^{2\phi })\partial_{\chi
}\,.}$$
 The corresponding finite transformation has three real invariants,
$$ e^{2\phi}-v^2 f^{-1}\equiv e^{2\phi_0}-v_0^2 f_0^{-1}, \quad
1-\beta =f^{-1}|\Phi|^2e^{2\phi}\equiv f_0^{-1}|\Phi_0|^2e^{2\phi_0},$$
$$\gamma=f^{-1}(\chi^2+\beta f^2)\equiv f_0^{-1}(\chi_0^2+\beta f_0^2),
\eqno (3.25)$$
and one complex one,
$$\nu=v+(if-\chi)\Phi^{-1}\equiv v_0+(if_0-\chi_0)\Phi_0^{-1},\eqno (3.26)
$$
 where $\Phi=u-zv$, and
$$ f=\chi \xi^{-1}=\gamma(\beta+\xi^2)^{-1}, \quad \xi=\chi_0 f_0^{-1}
-h_3 \gamma,$$
$$\Phi^{-1}=\Phi_0^{-1}+\nu h_3,\quad z=z_0-\nu^{-1}(\Phi-\Phi_0).\eqno
(3.27)$$
(In these formulas $h_1, h_2, h_3$ are the last real group parameters).

Altogether these Killing vectors form a closed ten-dimensional algebra.
Introducing instead of $K_s$ and $K_{d_3}$ two linear combinations
$K_1=(K_{d_3}-K_s)/2,\; K_2=-(K_{d_3}+K_s)/2,$ and enumerating  vectors
$K_m, K_e, K_g, K_{d_1}, K_{d_2}, K_{H_1}, K_{H_2}, K_E$ as
$K_3, K_4, \ldots, K_9, K_0$ respectively, we can write the commutation
relations
$$[K_a, K_b]={C^c}_{ab}K_c\,,\eqno (3.28)$$
where $a, b=0,1,...,9$, with the following nonzero structure constants:
$$\eqalignno{{C^3}_{13} &={C^5}_{15}={C^6}_{16}={C^4}_{24}=
{C^5}_{25}={C^7}_{27}={C^4}_{37}=\cr
 {C^8}_{30} &={C^3}_{46}={C^3}_{58}={C^8}_{69}={C^9}_{78}=
{C^1}_{67}=1,\cr
 {C^7}_{17} &={C^9}_{19}={C^0}_{10}={C^8}_{28}={C^6}_{26}=
{C^0}_{20}=\cr
 {C^4}_{59}&={C^9}_{40}={C^1}_{50}={C^2}_{50}={C^2}_{67}=-1,\cr
 {C^0}_{89}&={C^7}_{49}={C^6}_{38}=2,\;\;
   {C^2}_{48}={C^1}_{39}={C^5}_{34}=-2.&(3.29)}$$

This algebra is isomorphic to $so(2,3)$. To show this, consider a
five-dimensional pseudoeuclidean space,
$$d\sigma^2=G_{\mu\nu}d\xi^\mu d\xi^\nu=
-(d\xi^0)^2-(d\xi^\theta)^2+(d\xi^1)^2+(d\xi^2)^2+(d\xi^3)^2,\eqno (3.30)$$
and denote the corresponding $so(2,3)$ generators as $L_{\mu\nu}$,
where $\mu, \nu= 0, \theta, 1, 2, 3$. Then the following correspondence
between  $L_{\mu\nu}$ and the above Killing vectors can be found:
$$\eqalignno{ L_{01}&=\frac{1}{2}(K_{d_3}-K_s),\;\;
    L_{\theta 2}=-\frac{1}{2}(K_{d_3}+K_s),\;\;
    L_{03}=\frac{1}{2}(K_{H_2}-K_m),\cr
  L_{31}&=\frac{1}{2}(K_{H_2}+K_m),\;\;
     L_{32}=\frac{1}{2}(K_{H_1}+K_e),\;\;
    L_{\theta 3}=\frac{1}{2}(K_{H_1}-K_e),\cr
   L_{02}&=\frac{1}{2}(K_g-K_{d_1}-K_{d_2}-K_E),\;\;
     L_{1 \theta}=\frac{1}{2}(K_g+K_{d_1}+K_{d_2}-K_E),\cr
   L_{0\theta}&=\frac{1}{2}(K_g+K_{d_1}-K_{d_2}+K_E),\;\;
     L_{12}=\frac{1}{2}(K_g-K_{d_1}+K_{d_2}+K_E).&(3.31)}$$
Using (3.31) one can rewrite commutation relations (3.28) as
$$\eqalignno{[L_{\mu\nu}, L_{\lambda \tau}] &= G_{\mu \tau}L_{\nu \lambda}-
G_{\mu \lambda}L_{\nu \tau}+G_{\nu \lambda}L_{\mu \tau}-
G_{\nu \tau}L_{\mu \lambda}\cr
&= {C^{\alpha \beta}}_{\mu \nu\;\lambda \tau}
L_{\alpha \beta}\,.&(3.32)}$$

Different $so(1,2)$ subalgebras should correspond to $S$ and $T$
dualities. They can be chosen in various ways; physically this may
be traced to the possibility of gauge and scale transformations.
Indeed, one can see that both electric-magnetic duality and
Ehlers--Harrison generators enter into genuine $SO(2,3)$ generators
mixed with gauge and scale ones. Also it should be noted
that the standard procedure of Kaluza--Klein reduction does not
involve dualization of nondiagonal metric components and magnetic parts
of vectors, while our procedure does. Here we deal with the potential
space, as opposed to the metric space in the standard reduction. Both
should be related by a kind of Neugebauer--Kramer map, but we do not
enter into details of these matters here.

Now we are in a position to formulate the main result of the above
analysis: the target space (2.7) is a symmetric pseudoriemannian space
isomorphic to the coset $SO(2,3)/(SO(3)\times SO(2))$. Writing a
decomposition of the full algebra
$$\eqalignno{{\cal L} &=(
L_{0\theta},\;L_{12},\;L_{23},\;L_{13}),\cr
{\cal B}&=( L_{01},\;L_{02},\;L_{03},\;L_{\theta 1},\;
L_{\theta 2},\;L_{\theta 3}),&(3.33)}$$
one finds that
$$[{\cal B}{\cal B}]={\cal L},\quad
[{\cal B}{\cal L}]={\cal B},\quad
[{\cal L}{\cal L}]={\cal L}.\eqno (3.34)$$
This means that the homogeneous space generated by ${\cal B}$
is a symmetric space.
The equations of motion for $\varphi^A$ then are equivalent to the set
of conservation laws  for Noether currents
$$\partial_i(h^{ij}\sqrt{h}J_j^a)=0,\quad
J_i^a=\tau_A^a \;\frac{\partial\varphi^A}{\partial x^i}\,,\eqno (3.35)$$
built using the corresponding Killing one-forms
$${\bf\tau}^a =\eta^{ab}K^A_b {\cal G}_{AB}d\varphi^B,\eqno (3.36)$$
where  $\eta^{ab}$ is an inverse to the Killing--Cartan metric
$\eta_{ab}=k{C^c}_{ad}{C^d}_{bc}\,$.
This is a convenient form for further reduction to two dimensions.
\bigskip

\n {\bf 4. Stringy Kerr--NUT dyon}

Here we discuss an application of the target space isometries to
derive dilaton-axion counterparts to solutions of the vacuum
Einstein equations.
Any solution to the vacuum Einstein equations is a solution of the
present theory with $v=u=\kappa=\phi=0$. Therefore using the above
transformations
an axion-dilaton counterpart can be found to any stationary vacuum solution.
In this case the above formulas simplify considerably.  The first  Harrison
transformation will read
$$\frac{f}{f_0}=\frac{\chi}{\chi_0}=e^{2\phi}=\frac{1}{1-h_1^2 f_0}\,,\quad
v=h_1 f,\;\; u=h_1\chi,\;\; \kappa=h_1^2 \chi_0\,.\eqno (4.1)$$
If the seed solution is asymptotically flat, and one wishes
to preserve this property, it has to be accompanied by the scale
transformation (3.7) with the parameter
$e^{2s}=1-h_1^2$. The result can be
concisely expressed in terms of the Ernst potential ${\cal E}=f+i\chi$:
$${\cal E}=\frac{\sqrt{1-h_1^2}}{h_1}\left(v+iu\right)=
\frac{(1-h_1^2){\cal E}_0}{1-h_1^2 \hbox{ Re } {\cal E}_0}\,,\;\;\;
z=i\left(1-h_1^2{\cal E}_0 \right).\eqno (4.2)$$

A similar combined transformation via (3.22)--(3.7) reads
$${\cal E}=\frac{\sqrt{1-h_2^2}}{h_2}\left(u-iv\right)=
\frac{(1-h_2^2){\cal E}_0}{1-h_2^2 \hbox{ Re } {\cal E}_0}\,,\;\;\;
z = {i\over (1-h^2_2{\cal E}_0)}\,.\eqno (4.3)$$
In both cases  the metric rotation function is simply rescaled:\
$\omega_i=(1-h^2)^{-1} \omega_{0i}\,$,
where $h$ is either  $h_1$  or $h_2\,$.

When applied to vacuum solutions, our Ehlers-type transformation
reduces exactly to the original Ehlers transformation [19].
Indeed, in this case $\beta=1$ and from (3.27) we get
${\cal E}= {\cal E}_0(1+ih_3{\cal E}_0)^{-1}$,
while $v, u, \phi, \kappa$ remain zero.

Starting with the vacuum Kerr--NUT solution we have obtained [12]
a seven-parameter family of axion-dilaton black holes
using Harrison transformations in suitable combinations with
other isometries.
The resulting metric may be written in the same form as the vacuum
Kerr--NUT solution,

$$ds^2=\frac{\Delta-a^2 \sin^2\theta}{\Sigma}
\left(dt-\omega d\varphi\right)^2
-\Sigma\left(\frac{dr^2}{\Delta}+d\theta^2+
\frac{\Delta\sin^2\theta}{\Delta-a^2\sin^2\theta} d\varphi^2\right),
\eqno (4.4)$$
with somewhat modified characteristic functions
$$\eqalignno{ \Delta &=(r-r_-)(r-2M) + a^2 - (N-N_-)^2,\cr
\Sigma &= r(r-r_-)+(a\cos\theta+N)^2 -N_-^2,&(4.5)\cr
\omega &=\frac{2}{a^2\sin^2\theta-\Delta}
\left\{N\Delta\cos\theta+a\sin^2\theta
\left[M(r-r_-)+N(N-N_-)\right]\right\}. }$$
The corresponding electric and magnetic potentials and the axidilaton
field are
$$\eqalignno{v &= \frac{\sqrt{2}e^{\phi_{\infty}}}{\Sigma}
\hbox{ Re} \left[{\cal Q} (r-r_-+i\delta)\right],\quad
u=\frac{\sqrt{2}e^{\phi_{\infty}}}{\Sigma} \hbox{ Re}\left[{\cal Q}
z_{\infty}(r-r_-+i\delta)\right],&(4.6)\cr
z &= \frac{z_{\infty}\rho+{\cal D}z^*_{\infty}}{\rho+{\cal D}}\,,\quad
\rho=r -\frac{{\cal M^*}r_-}{2M}+i\delta,\; \delta=a\cos\theta +N-N_-\,.
&(4.7)}$$
Here ${\cal M}=M+iN$ is the complex mass, ${\cal Q}=Q-iP$ is the complex
charge. A complex dilaton-axion charge ${\cal D}=D+iA$,
defined asymptotically as
$$z=z_{\infty}-2ie^{-2\phi_{\infty}} \frac{{\cal D}}{r} +
O\left(\frac{1}{r^2}\right),\eqno (4.8)$$
is related to the other parameters as follows:
$${\cal D}=-\frac{{\cal Q}^{*2}}{2{\cal M}}.\eqno (4.9)$$
Note that this relation does not involve the rotation parameter $a$.
Separately dilaton and axion charges are
$$D=\frac{M(P^2-Q^2)-2PQN}{2(M^2+N^2)}\,,\quad
A=\frac{N(Q^2-P^2)-2PQM}{2(M^2+N^2)}\,.\eqno (4.10)$$
Two additional parameters in the metric functions read
$$r_-=\frac{M|{\cal Q}|^2}{|{\cal M}|^2}\,,\quad N_-=\frac{N|{\cal Q}|^2}
{2|{\cal M}|^2}\,.\eqno (4.11)$$

Hence the solution contains seven independent real parameters:
a mass $M$, a rotation parameter $a$, a NUT--parameter $N$,
electric $Q$ and
magnetic $P$ charges (defined as in [20] to have
the standard asymptotic normalization of the Coulomb energy),
and asymptotic values of the axion $\kappa_{\infty}$
and the dilaton   $\phi_{\infty}$ (combined in  $z_{\infty}$). The
complex axidilaton charge introduced through an asymptotic expansion
 is determined by the electromagnetic charge and the complex mass:

This family containes as particular cases many previously known
solutions to dilaton-axion gravity. For $N=P=0$ it corresponds to
Sen's solution [21] up to some coordinate transformation
(in this case the axion charge $A=0$). For $a=0$
it coincides (up to a transformation
of the radial coordinate) with the 6-parameter solution
reported recently by Kallosh {\it et al.\/} [20]; its
3-parameter subfamily was also found by  Johnson and Myers [22].
For $N=0,\; a=0$ we recover the 5-parameter
solution presented by Kallosh and Ortin [23], and
if in addition one of the charges $Q, P$ is zero, the solution
reduces to the Gibbons--Maeda--Garfinkle--Horowitz--Strominger black hole
[16]. Finally, when $P=Q=0$ we come back to the Kerr--NUT metric.

As in vacuum and electrovacuum cases,
for $N\neq 0$ our solution cannot be
properly interpreted as a black hole because of time periodicity,
which is to be imposed in the presence of the wire singularity.
We can still use the notation $r_H^{\pm}$ for the values
of the radial coordinate marking  positions of the surfaces where
$\Delta=0$:
$$r_H^{\pm}=M+r_-/2\pm\sqrt{|{\cal M}|^2(1-r_-/2M)^2-a^2}.\eqno (4.12)$$
For $N=0$ the upper value $r_H^+$ corresponds to the event horizon of a
black hole. The timelike Killing vector $\partial_t$ becomes null at the
surface $r=r_e(\theta)$,
$$r_e^{\pm}=M+r_-/2\pm\sqrt{|{\cal M}|^2(1-r_-/2M)^2-a^2\cos^2\theta},
\eqno (4.13)$$
which marks the boundary of a black hole ergosphere in the case  $N=0$.
Inside the 2-surface  $r=r_e(\theta)$ the Killing vector
$\partial_t-\Omega \partial_\varphi$ with some $\Omega= \hbox{const}\,$ may
still be timelike, the boundary
value of $\Omega$ at $r=r_H^+$ where it becomes null being
$$\Omega_H=\frac{a}{2}\left\{|{\cal M}|^2(1-r_-/2M)+M
\sqrt{|{\cal M}|^2(1-r_-/2M)^2-a^2}\right\}^{-1}.\eqno (4.14)$$
For $N=0$ this quantity has a meaning of the angular velocity of the horizon.
The area of the two--surface  $r=r_H^+$ is
$$A=4\pi a/\Omega_H.\eqno (4.15)$$

The square root in (4.12) becomes zero for the family of extremal solutions.
This corresponds to the  relation between the parameters
$$|{\cal D}|=|{\cal M}| -a,\eqno (4.16)$$
which defines a 4-dimensional hypersurface in the 5-dimensional
space of parameters $Q, P, M, N, a$.
For extremal solutions we have
$$\eqalignno{
r_H^{\rm ext} &= 2M-\frac{aM}{|{\cal M}|}\,,\quad
\Delta^{\rm ext}=\left(r-r_H^{\rm ext}\right)^2,\cr
\omega^{\rm ext} &=\frac{2\left\{N \Delta^{\rm ext}\cos\theta +
a\sin^2\theta \left[M\left(r-r_H^{\rm ext}\right)+a|{\cal
M}|\right]\right\}} {a^2\sin^2\theta-\Delta^{\rm ext}}\,, &(4.17)\cr
\Sigma^{\rm ext} &=2M\left(r-r_H^{\rm ext}\right)+
\Delta^{\rm ext}-a^2\sin^2\theta +2a\left(|{\cal M}|+N\cos\theta\right).}$$

The metric for the {\it non--rotating} extremal dilaton-axion
Taub--NUT family reads
$$\eqalignno{ds^2 &=\left(1-2M/r\right) \left(dt+2N\cos\theta d\varphi\right)^2-
\left(1-2M/r\right)^{-1} dr^2\cr
&\quad  -r(r-2M)\left(d\theta^2+\sin^2\theta
d\varphi^2\right).&(4.18)}$$
In this case $r_H^{\rm ext}=2M$,  coinciding with the curvature
singularity. (Note that this is not so if $a\neq 0$, since then
$\Sigma^{ext}(r_H^{ext})\neq 0$.) For the dilaton we get from (4.7)
$$e^{2\left(\phi-\phi_{\infty}\right)}=\frac{1}{r(r-2M)}
\left|r-2M+\frac{i(QN-PM){\cal Q}^*}{|{\cal M}|^2}\right|^2.\eqno (4.19)$$
Comparing (4.18) and (4.19) one can see that generally the string metric
 $ds^2_{\rm string}=e^{2\phi} ds^2$
has nonsingular throat structure. However, if
$$QN=PM,\eqno (4.20)$$
the dilaton factor (4.19) vanishes as $r\rightarrow 2M$ and the
string metric will have the same behavior as in the case
of the static dilaton
electrically charged black hole [16] (to which our solution
reduces if $N=P=0$). Hence, regular Taub--NUT string throats
form a 3-parameter family corresponding to the hypersurface
$|{\cal D}|=|{\cal M}|$
in the parameter space of $M, N, P, Q$, from which a 2-dimensional subspace
(4.20) has to be excluded. As  was shown recently by Johnson [24],
some of the family of extremal Taub--NUT solutions
have exact gauged WZW model counterparts.\bigskip

\n {\bf 5. Two-dimensional integrability}

 If the target space of a three-dimensional sigma model is a {\it
symmetric} riemannian space $G/H$
with $N$--parameter isometry group $G$ acting transitively on it ($H$ being
an isotropy subgroup), generated  by the set of $N$
Killing vectors forming the Lie algebra of $G,\;\;[{K_a, K_b }]
={C^c}_{ab}K_c$,  $a, b, c=1,...,N,$
the current conservation equations (3.35) may be cast into the null-curvature
form. With a proper normalization of the Killing--Cartan metric on the
isometry group  the one-forms dual to the Killing vectors
will satisfy the Maurer--Cartan equation with the same structure constants,
$$d{\bf\tau}^a +\frac{1}{2}{C^a}_{bc}{\bf\tau}^a\wedge
{\bf\tau}^b=0.\eqno (5.1)$$
Let $e_a$ denote some matrix
representation of the Lie algebra of
$G,\; [e_a,e_b]={C^c}_{ab}e_c\,$.
Define the following matrix-valued connection one-form:
${\cal A}={\cal A}_Bd \varphi^B = e_a {\bf\tau}^a$.
In view of (5.1), the corresponding curvature vanishes,
$${\cal F}_{BC}={\cal A}_{C,B}-{\cal A}_{B,C}+ [{\cal A}_B,{\cal A}_C]=0,
\eqno (5.2)$$
and thus ${\cal A}_B$ is  pure gauge,
$${\cal A}_B= -(\partial_B g) g^{-1},\quad g\in G .\eqno (5.3)$$
Because of the gauge invariance under $H$, the matrix $g$ actually belongs
to the coset $G/H$ [6]. The pullback of ${\cal A}$
onto the configuration space ${x^i}$ is
equivalent to (3.35) and, hence, to the equations of
motion of the sigma model. In terms of $g$ the equations (3.35) read
$$d\{(\star dg)g^{-1}\}=0,\eqno (5.4)$$
where a star stands for a 3-dimensional Hodge dual.

Now reduce the system to two dimensions, imposing an axial symmetry
condition. In order to ensure regularity on the polar axis,
hypersurface orthogonality has to be supposed in addition. Then the 3-metric
may be conveniently written in the Lewis--Papapetrou gauge:
$$h_{ij}dx^idx^j=e^{2\gamma}(d\rho^2 + dz^2)+\rho^2 d\varphi^2,\eqno
(5.5)$$
and (5.4) becomes equivalent to a modified chiral equation
$$(\rho g_{,\rho} g^{-1})_{,\rho} +  (\rho g_{,z} g^{-1})_{,z}=0.\eqno
(5.6)$$
A variety of techniques is available now to deal with such systems.
A Lax pair with the complex spectral parameter $\lambda$ can be readily
given (we use the Belinskii--Zakharov form [1]):
$$D_1\Psi=\frac{\rho U-\lambda V}{\rho^2+\lambda^2}\Psi,\quad
D_2\Psi=\frac{\rho V+\lambda U}{\rho^2+\lambda^2}\Psi,\eqno (5.7)$$
where $V=\rho g_{,\rho}g^{-1},\; U=\rho g_{,z}g^{-1}, \; \Psi$
is a matrix ``wave function"\negthinspace, and
$$D_1=\partial_z-\frac{2\lambda^2}{\rho^2+\lambda^2}\partial_{\lambda}\,,
\quad
D_2=\partial_{\rho}+\frac{2\lambda \rho}{\rho^2+\lambda^2}\partial_{\lambda}
\eqno (5.8)$$
are commuting operators. A chiral system (5.6) follows from the compatibility
condition $[D_1, D_2]\Psi=0$.
This linearization is sufficient to establish a desired integrability
property. The inverse scattering transform method
can be directly applied to (5.7) to generate multisoliton
solutions, and an infinite-dimensional Kac--Moody algebra
can be derived [6].

In view of the results of Sec.\ 3, the EMDA system in two dimensions
will be described by the chiral model based on the coset
$ SO(3,2)/(SO(3)\times SO(2))$ [25].
It is worth noting that for $D=4,\;p=1$ theory
$T$-duality in {\it two} dimensions is $SO(2,3)$---i.e., the same as an
enhanced
symmetry found here in {\it three} dimensions. However it is just because
of existence of the same symmetry in three dimensions (ensuring the
symmetric-space property of the target space) that the corresponding
two-dimensional
symmetry becomes infinite and is generated by the $SO(2,3)$ current algebra.
A similar situation was recently described by Bakas [26] in the
dilaton-axion gravity without vector field ($p=0$), where an analogous
role was played by the group $SO(2,2)$.

Let us describe the one-form representation of the present theory
more explicitly [25].
Using the pair-index notation, the Killing--Cartan metric
of $so(2,3)$ can be written as
$$\eta_{\mu\nu\;\lambda\tau}=
1/12\; {C^{\alpha\beta}}_{\mu\nu\;\gamma\delta}
{C^{\gamma\delta}}_{\lambda\tau\;\alpha\beta}\,,\eqno (5.9)$$
and the target space metric in terms of the desired one--forms reads
$${\cal G}_{AB}=
1/2\; \eta_{\mu\nu\;\lambda\tau} \tau^{\mu\nu}_A \tau^{\lambda\tau}_B.
\eqno (5.10)$$
Using Killing vectors $K_a$ and the correspondence rules (3.31) one
can construct $\tau$--forms (3.36). The Abelian subalgebra will read:
$$\tau^{01}=-(\omega_1 +\omega_f),\;\;\quad \tau^{\theta 2}=\omega_1
-\omega_f+2\omega_2\,, \eqno (5.11)$$
where
$$\eqalignno{\omega_1 &=\kappa
\omega_{\kappa}-2d\phi+u(v\omega_{\chi}-2\omega_u),\cr
\omega_f &=f^{-1}df+\chi \omega_{\chi}\,,\;\;\quad
\omega_2=v\omega_v+w\omega_u\,,\cr
\omega_{\kappa} &=e^{4\phi}
d\kappa,\;\;\quad \omega_{\chi}=f^{-2}(d\chi+vdu-udv),\cr
 \omega_v& =f^{-1}e^{-2\phi} dv,\;\quad
\omega_u=f^{-1}e^{2\phi} (du - \kappa dv).&(5.12)}$$
The other components look more complicated:
$$\eqalignno{2\tau^{0\theta}
&=\omega+\omega_6-\omega_7-\omega_{\chi}\,,\;\;\quad
2\tau^{02}=\omega-\omega_6-\omega_7+\omega_{\chi}\,,\cr
-2\tau^{\theta 1}&=\omega+\omega_6+\omega_7+\omega_{\chi}\,,\;\;\quad
2\tau^{12}=\omega-\omega_6+\omega_7-\omega_{\chi}\,,&(5.13)\cr
-\tau^{03} &=\omega_5+\omega_8,\;\quad
\tau^{13}=\omega_5-\omega_8\,,\;\quad
\tau^{\theta 3}=\omega_4-\omega_9,\;\quad
-\tau^{23}=\omega_4+\omega_9\,,}$$
where the following recurrent sequence is used:
$$\eqalignno{\omega_3 &= \kappa \omega_u - \omega_v\,,\;\quad
\omega_4=u\omega_{\chi} - \omega_3\,,\;\quad
\omega_5=v\omega_{\chi} - \omega_u\,, \cr
 \omega_6 &=d\kappa -\kappa^2\omega_{\kappa} +4\kappa d\phi -
u(\omega_4 - \omega_3),\;\;\quad
\omega_7=\omega_{\kappa} +v( \omega_5-\omega_u),\cr
\omega_8 &=u\tau^{01} -v\omega_6+\chi \omega_3 -u\omega_2+du,&(5.14)\cr
\omega_9 &=v\tau^{\theta 2} -u\omega_7+\chi \omega_u -v\omega_2+dv,\cr
\omega &=u\omega_9 -v\omega_8+\chi(\chi\omega_{\chi}+\omega_2-2\omega_f)
+d\chi.}$$

Now, given an adjoint matrix representation of $so(3,2)$, one can build
a $5\times 5$ connection one-form ${\cal A}$ and the corresponding matrix
$g \in SO(3,2)/(SO(3)\times SO(2))$. Remarkably, the present theory
also admits a more concise representation in terms of the symplectic
$4\times 4$ matrices (due to the isomorphism $SO(3,2)\sim Sp(4,R)$).
The symplectic connection can be written in block form using three
real $2\times 2$ matrices, two of which ($B,\;D$) are symmetric,
$$\eqalignno{ B &=\frac{1}{2}\left\{ (\tau^{0
\theta} - \tau^{\theta 3})I_2 +
 (\tau^{23} - \tau^{02})\sigma_x +
 (\tau^{01} - \tau^{13})\sigma_z \right\},\cr
 C &=\frac{1}{2}\left\{\tau^{03}I_2 -\tau^{\theta 2}\sigma_x-
i\tau^{12}\sigma_y+\tau^{\theta 1}\sigma_z\right\},&(5.15)\cr
  D&= \frac{1}{2}\left\{ -(\tau^{0\theta} + \tau^{\theta 3})I_2 -
 (\tau^{23} + \tau^{02})\sigma_x +
 (\tau^{01} + \tau^{1 3})\sigma_z \right\},}$$
as follows:
$${\cal A}=\left(
\matrix{C & B \cr
D & -C^T \cr}
\right).\eqno (5.16)$$
Here $I_2$ is a unit matrix and $\sigma_x, \sigma_y, \sigma_z$ are Pauli
matrices with $\sigma_z$ diagonal. In view of (5.4), the equations of motion
of the EMDA sigma model are equivalent to vanishing of the curvature
(5.2) related to (5.16). This implies the existence of the symmetric symplectic
$4\times 4$ matrix $g \in Sp(4,R)/U(2)$ entering the Belinskii--Zakharov
representation (an explicit form will be given elsewhere).\bigskip

\n {\bf 6. Dilaton gravity: critical couplings}

Now consider the target space (2.9) of the purely dilatonic model.
Of the ten Killing vectors of the dilaton-axion theory, four
($K_g, \;K_e,\; K_m,\; K_s$) are independent of the axion; they remain
isometry generators for the dilatonic model too. However, $sl(2,R)$
duality is broken to a dilaton shift (inherited from $K_{d_3}$):
$$K_\phi=v\partial_v - a \partial_a + \alpha^{-1} \partial_{\phi}\,.\eqno
(6.1)$$
(When $\alpha \rightarrow 0$  it should be renormalized by multiplying  by
$\alpha$; in this limit $K_\phi$ becomes a pure dilaton shift.)
The corresponding finite transformation is
$$v\rightarrow e^d v,\quad u\rightarrow e^{-d}u,\quad \phi
\rightarrow \phi + \alpha^{-1} d,\eqno (6.2)$$
 where $d$ is real parameter.
These five generators form a closed algebra. Enumerating them
>from 1 to 5 as in Sec.\ 3 (with $K_\phi$ instead of $K_{d_3}$) we
get the following nonzero structure constants:
$${C^3}_{13}={C^4}_{24}={C^5}_{15}={C^5}_{25}= 1,\;\; {C^5}_{34}=-2.
\eqno (6.3)$$
This algebra is solvable.  Denoting it as ${\cal K}$ one can see that its
derivative ${\cal K}'$ contains as basis
vectors $K_3, K_4, K_5$;  the second derivative is
one-dimensional:\ ${\cal K}^{''}=K_5\,$; and we have the  chain of
subalgebras
$$0={\cal K}^{'''}\subset {\cal K}^{''}\subset {\cal K}^{'}\subset {\cal
K}, \eqno  (6.4)$$
where each term is an ideal of the preceding one.
It can  also be shown that the Killing--Cartan bilinear form
is degenerate. Since the target space is five-dimensional, this means that
the above symmetries are insufficient for creating a
symmetric-space structure.

Such algebras are known to admit a representation in terms of
 upper-triangular matrices.  Consistently with (6.3) $K_1$ and $K_2$ can be
choosen diagonal.  Then the following 3$\times$3 representation
$K_a\rightarrow e_\mu$ holds:
$$\eqalignno{
e_1 &=\frac{1}{3} \left(\matrix{
2 & 0 & 0\cr
0 & -1 & 0\cr
0 & 0 & -1\cr}
\right),\quad
e_2=\frac{1}{3} \left(\matrix{
1 & 0 & 0\cr
0 & 1 & 0\cr
0 & 0 & -2\cr}
\right),\quad
e_3=\left(\matrix{
0 & 1 & 0\cr
0 & 0 & 0\cr
0 & 0 & 0\cr
}\right),\cr
e_4 &=\left(\matrix{
0 & 0 & 0\cr
0 & 0 & -1\cr
0 & 0 & 0\cr}
\right),\quad
e_5=\frac{1}{2}\left(\matrix{
0 & 0 & 1\cr
0 & 0 & 0\cr
0 & 0 & 0\cr}
 \right).&(6.5)}$$
Obviously, this set constitutes  a basis for the upper-triangular
subalgebra of $sl(3,R).$

It is instructive to carry out a further geometric analysis of the
situation.
Computing the Riemann tensor, one can find that both the scalar curvature,
$$R=-(\alpha^2 +12),\eqno (6.6)$$
and the square of the Riemann tensor,
$$R_{ABCD}R^{ABCD}=3(\alpha^4 +16),\eqno (6.7)$$
are coordinate-independent. However, for general $\alpha$
the target space is not a symmetric space. Direct computation shows that
almost all covariant derivatives of the curvature
tensor are zero, except for the following:
$$R_{vuv\phi ;u}=R_{vuu\phi ;v} = -\,\frac{\alpha(\alpha^2
-3)}{f^2}\,.\eqno (6.8)$$
This means the target space (2.9) is symmetric only for
$\alpha=0, \pm \sqrt{3}$ [27].
Since sign of $\alpha$ is irrelevant (it is compensated by the reflection
of the dilaton $\phi\rightarrow -\phi$ in (2.8)) we conclude that only BMD and
KK theories are exceptional within the class under consideration.

Also, it turns out that the target space is an ``almost'' Einstein space.
Indeed,
$$R_{AB}=-3{\cal G}_{AB}\eqno (6.9)$$
for $A,B=f, \chi, v, u, \phi \; ({\cal G}_{AB}$ 
is the target space metric), but
$$R_{\phi \phi}=-\alpha^2 {\cal G}_{\phi \phi}\,.\eqno (6.10)$$
For $\alpha^2=3$ the target space is both a homogeneous symmetric space
and an Einstein space.

Let us consider two critical cases in more detail.
For $\alpha=0$  the present theory coincides with
the BDM model with the Brans--Dicke parameter $\omega =-1$.
The presence of the scalar field just adds a trivial generator
(constant dilaton shift) to the $su(2,1)$ algebra
of the coresponding EM theory. The latter includes four Killing vectors
found above ($K_1, K_2, K_3, K_5$) and a continuous electric-magnetic
duality rotation
$$K_d= (v\partial_u - u\partial_v),\eqno (6.11)$$
which is broken if $\alpha \neq 0$.

The nontrivial part of the 8-parameter $su(2,1)$ symmetry algebra
consists of Harrison
transformations [15]
$$\eqalignno{
K_{H_1} &= 2fv\partial_f + \left[v\chi +uf -u(u^2 +
v^2)/2\right]\partial_{\chi}  \cr
&\quad +\left[(v^2-3u^2)/2 +f\right]\partial_v + (2uv+\chi)\partial_u,
&(6.12)\cr
K_{H_2} &= 2fu\partial_f + \left[u\chi -vf +v(u^2 +v^2)/2\right]
\partial_{\chi}  \cr
&\quad +\left[(u^2-3v^2)/2 +f\right]\partial_u + (2uv-\chi)\partial_v,
&(6.13)}$$
and the EM version of the Ehlers transformation
generated by the commutator of the Killing vectors (see (3.23))
$$K_E= 2f\chi\partial_f + (\chi^2 -F^2) \partial_{\chi} +
(\chi v+uF)\partial_v + (\chi u-vF)\partial_u,\eqno (6.14)$$
where $F=f-(v^2+u^2)/2$.
Already from the infinitesimal form it can be seen that Harrison
transformations mix gravitational and electromagnetic potentials,
while Ehlers transformations mix gravitational variables.
Unfortunately, these symmetries turn out to be broken by the dilaton.

In the case $\alpha=\sqrt{3}$ (KK-theory)
one encounters a remarkable counterpart to Harrison transformations.
The infinitesimal generators  were found by Neugebauer in 1969 [18].
In our notation Neugebauer transformations read
$$\eqalignno{
K_{N_1} &= 2fv\partial_f + (v\chi +ufe^{2\alpha \phi}+uv^2) \partial_{\chi}
 \cr
&\quad +(2v^2+f e^{2\alpha \phi})\partial_v - (uv-\chi)\partial_u
+\sqrt{3}v \partial_{\phi}\,,&(6.15)\cr
K_{N_2} &= 2fu\partial_f + (u\chi -vfe^{-2\alpha \phi}- u^2v)
\partial_{\chi}  \cr
&\quad +(2u^2+fe^{-2\alpha \phi})\partial_u - (uv+\chi)\partial_v
-\sqrt{3}u \partial_{\phi}\,.&(6.16)}$$
The commutator of
two Neugebauer transformations gives a KK analog of the Ehlers
transformation,
 $$[K_{N_1},K_{N_2}]=2 K_{EN},\eqno (6.17)$$
which explicitly reads
$$\eqalignno{
&K_{EN} = 2f\chi\partial_f + \left[\chi^2 -f^2+f\left(v^2e^{-2\alpha \phi}+
u^2e^{2\alpha \phi}\right)+u^2v^2\right] \partial_{\chi} \cr
&\quad +\left[v(uv+\chi)+ufe^{2\alpha \phi}\right]\partial_v
 +\left[u(\chi-uv)-vfe^{-2\alpha \phi}\right]\partial_u +\sqrt{3}vu
\partial_{\phi}\,.&(6.18)}$$

In spite of an apparent similarity of the Ehlers--Harrison transformations
for $\alpha=0$ and the Neugebauer transformations for $\alpha=\sqrt{3}$, no
continuous $\alpha$-interpolation can be found between them within
the EMD theory.
\bigskip

\n {\bf 7. Diagonal metrics}

If the rotation one-form $\omega_i$ in (2.1) is zero and, in addition,
either the electric potential $v$ or the magnetic  potential $u$ is set to
zero too, the axion decouples in the EMDA system and hence can consistently
be chosen zero. Both EMDA and EMD theories are then cast into
electrostatics and magnetostatics, and in this situation the EMD theory is
more general since the dilaton coupling constant is arbitrary. So we take
as a starting point the target space (2.9). Setting
$\chi=0=u$ we are left with the three-dimensional space
$$dl^2_e=\frac{df^2}{2f^2}-\frac{e^{-2\alpha \phi}dv^2}{f}+2d\phi^2.
\eqno (7.1)$$
 It can be checked that all covariant derivatives of the corresponding
Riemann tensor vanish, i.e., we are dealing  with a symmetric space.
It is, however, not an Einstein space. The nonzero components of the Ricci
tensor read
$$R_{ff}=-\,\frac{1}{4f^2}\,,\;\; R_{f\phi}=\,\frac{\alpha}{2f}\,,\;\;
R_{vv}=\frac{\nu e^{-2\alpha \phi}}{f}\,,\;\;R_{\phi\phi}=-\alpha^2,
\eqno (7.2)$$
where $\nu=(\alpha^2+1)/2$. Clearly $R_{AB}$ is not proportional
to the metric tensor in (7.1) except for $\alpha=0$, if we set in addition
$\phi\equiv 0$.

Solving the Killing equations for the metric (7.1) one finds four Killing
vectors:
$$\eqalignno{K_0 &=\frac{1}{2}\partial_\phi -\alpha f \partial_f,\quad
K_1=\partial_v, \cr
K_2 &=\frac{1}{2}\left(v^2 +\nu^{-1}f e^{2\alpha\phi}\right)\partial_v +
\frac{v}{2\nu}\left(\alpha \partial_\phi +2f\partial_f\right),&(7.3)\cr
K_3 &=v\partial_v+\nu^{-1}\left(\frac{\alpha}{2}\partial_\phi+
f\partial_f\right).}$$
The first one commutes with all others:
$$\left[K_0,K_i\right]=0,\;\;\; i=1, 2, 3,\eqno (7.4)$$
while the $K_i$ form an algebra $sl(2,R)$:
$$ \left[K_1,K_2\right]=K_3,\quad
 \left[K_1,K_3\right]=K_1,\quad
 \left[K_2,K_3\right]=-K_2.\eqno (7.5)$$
Here the Harrison-type generator is $K_2$, and it is worth noting that for
$\alpha=\sqrt{3}$ the corresponding generator $K_{N_1}$ (6.15) does 
reduce to $K_2$ restricted to the static subspace.

Finite transformations corresponding to these infinitesimal symmetries
can be found by a straightforward integration.
Consider first an electrostatic case. The vector $K_0$ generates the
 transformation
$$\phi\rightarrow \phi+\frac{\lambda_0}{2}\,,\quad
f\rightarrow fe^{-\alpha\lambda_0},\quad
v\rightarrow v ,\eqno (7.6)$$
where $\lambda_0$ is a real parameter.
It can be interpreted as a constant dilaton shift, accompanied by
rescaling of the three-dimensional conformal factor. Comparing with
the results of the Sec.\ 3 one  sees that it is a superposition of the
stationary dilaton shift (6.1) and the scale transformation (3.8)
restricted to $\chi\equiv 0,\; u\equiv 0$.

The second Killing  vector $K_1$ obviously corresponds to a gauge
transformation of the electrostatic potential,
$$ v\rightarrow v +\lambda_1,\quad
 f\rightarrow f,\quad \phi\rightarrow \phi,\eqno (7.7)$$
and is directly related to the stationary gauge transformation (3.3).

An essentially nontrivial (Harrison-type) transformation is generated
by $K_2$.  Integration  gives an invariant:
$$fe^{-2\phi/\alpha}=f_0e^{-2\phi_0/\alpha},\eqno (7.8)$$
and a linear transformation law for the quantities
$$s=\frac{v e^{-\alpha\phi}}{\sqrt{f}}\,,\quad
 t=\frac{\sqrt{f} e^{\alpha\phi}}{\nu} -\frac{v^2
e^{-\alpha\phi}}{\sqrt{f}}; \eqno (7.9)$$
$$t=t_0,\quad s=s_0+t_0\lambda_2/2,\eqno (7.10)$$
where the index $0$ corresponds to $\lambda_2=0$.
Finally, equations generated by $K_3$ give the following last
isometry of the target space (7.1):
$$v\rightarrow ve^{\lambda_3},\quad  f\rightarrow fe^{\lambda_3/\nu},\quad
\phi \rightarrow  \phi + \frac{\alpha\lambda_3}{2\nu}\,.\eqno (7.11)$$

To select  combinations of these transformations which
preserve  asymptotic flatness it is convenient to write down the general
 four-parameter isometry transformation obtained by successive
application of the above transformations:
$$
f=f_0 \Lambda^{-1/\nu} e^{-\alpha\lambda_0 +\lambda_3/\nu}, $$
$$\exp (\phi) =\Lambda^{-\alpha/(2\nu)} \exp \left(\phi_0+\lambda_0/2
+\alpha\lambda_3/(2\nu)\right),\eqno (7.12)$$
$$v=e^{\lambda_3}\left( v_0+
\frac{1}{2}t_0 \lambda_2 \sqrt{f_0}e^{\alpha\phi_0}\right)\Lambda^{-1}+
\lambda_1, $$
where
$$\Lambda=\frac{1-\nu s^2}{1-\nu s_0^2}\,.\eqno (7.13)$$
It can be shown that
the Harrison-like transformation in dilaton gravity  recently used for
generating purposes  [14], [17]
is a particular case of this four-parameter class.

Magnetostatics ($\chi=0=v$),
$$dl^2_m=\frac{df^2}{2f^2}-\frac{e^{2\alpha \phi}du^2}{f}+2d\phi^2,
\eqno (7.14)$$
can be treated along the same lines. Moreover, by reparametrization
$$\xi=(\alpha\phi-1/2\;\ln f)/\mu,\;\;\quad \eta=(\phi+\alpha/2\; \ln
f)/\mu, \eqno (7.15)$$
for the  magnetic case, and
$$\xi=-(\alpha\phi+1/2\;\ln f)/\mu,\;\;\quad \eta=(\phi-\alpha/2\; \ln
f)/\mu, \eqno (7.16)$$
for the electric one, where $\mu^2=\nu$, one gets a unified
description of both cases.  Denoting by $u$ either the magnetic ($u$)
or the electric ($v$) potential, one can represent
the line element of the truncated target space as
$dl_3^2=d\eta^2+dl_2^2$, where
$$dl_2^2=d\xi^2-e^{2\mu\xi} du^2.\eqno (7.17)$$
This 2-dimensional  space can  be easily shown to
represent a coset $SL(2, R)/U(1)$. Indeed,
one can find three Killing vectors for (7.17):
$$K_1=\partial_u,\;\; K_2=p\partial_u-\mu^{-1}u \partial_{\xi},\;\;
K_3=u\partial_u-\mu^{-1}\partial_\xi,\eqno (7.18)$$
where $p=(u^2+\mu^{-2}e^{-2\mu\xi})/2$,
with the $sl(2, R)$
structure constants ${C^3}_{12}={C^2}_{32}={C^1}_{13}=1$.
 The corresponding Killing--Cartan
one-forms, with the normalization $k=(2\mu)^{-2}$, will satisfy 
(5.1), and we have
$dl^2_2=1/2\; \eta_{ab}{\bf\tau}^a\otimes{\bf\tau}^b$, where
$\eta_{ab}=2k \;{\rm diag} (1,1,-1)$.
Choosing as $e_a$ a $2\times 2$ representation of $sl(2, R)$,
one can find using (5.3) the following  matrix $g \in SL(2,R)/U(1)$:
$$g=\mu e^{\mu\xi}\sqrt{2}\left(\matrix{
u^2-p & -u/\sqrt{2} \cr
-u/\sqrt{2} & 1\cr}
\right)\eqno (7.19)$$
which can be used in the Lax-pair (5.7) in the axisymmetric case.

Perhaps, all these considerations were unnecessary in order
to reveal the two-dimensional integrability of the arbitrary-$\alpha$
EMD theory in the diagonal case. Indeed, for $\alpha=0$ ($\nu=1/2$)
the EMD theory reduces to the
corresponding representation for the electrovacuum. Since, as we have
shown, the
underlying algebraic structure is $\alpha$-independent, this fact is
  already sufficient to reveal integrability of the static axisymmetric EMD
system
with arbitrary $\alpha$. However, an explicit construction may be useful
for practical generating purposes. Note that, contrary to the stationary EMDA
case, here we can extend any diagonal two-Killing solution of the EM theory
to the arbitrary-$\alpha$ EMD theory.\bigskip

\n {\bf 8. Conclusions}

We have shown that the EMDA theory reduced to three dimensions possesses
a ten-parameter global symmetry group $SO(2,3)$ containing
$T$ and $S$ dualities as different $SO(1,2)$ subgroups.
This theory may be represented as a sigma model on the symmetric
coset space $SO(2,3)/(SO(3)\times SO(2))$. Therefore,
further two-dimensional reduction gives an integrable system with
associated infinite symmetries. These features were found in the
$D=4,\; p=1$ model, but it is likely that they will persist as well for
the full ten-dimensional bosonic heterotic string effective
action
\footnote{$^1$}{{\sl Note added (July 3rd, 95):\/} After this talk was
given, a number
of papers appeared closely related to the subject. An enhancement of the
heterotic string symmetries in three dimensions was studied for the
full effective action by A. Sen ({\it Nucl. Phys.} {\bf B434}
(1995) 179); more general aspects were discussed by C. Hull
and P. Townsend ({\it Unity of Superstring
Dualities}, hep-th/9410167). Reduction to two dimensions was discussed
by J. Maharana ({\it Hidden Symmetries of Two Dimensional
String Effective Action}, hep-th/9502001,
{\it Symmetries of the Dimensionally Reduced
String Effective Action}, hep-th/9502002), A. Sen ({\it Duality
Symmetry Group of Two Dimensional Heterotic String Theory,
}, TIFR-TH-95-10, hep-th/95030157) and J.H. Schwarz
({\it Classical Symmetries of Some
Two-Dimensional Models}, preprint CALT-68-1978, hep-th/9503078;
{\it Classical Duality Symmetries in Two Dimensions},
preprint CALT 68--1994, hep-th/9505170). Related discussion can be
found also in
A. Kumar, K. Ray, {\it  Ehlers Transformations and String Effective Action},
preprint IP/BBSR/95-18, hep-th/9503154,
A.K. Biswas, A. Kumar, K. Ray, {\it Symmetries of Heterotic String Theory},
preprint IP/BBSR/95-51, hep-th/9506037,
I.R. Pinkstone, {\it Structure of dualities in bosonic string theory},
DAMTP R94-62, hep-th/9505147. The Kerr--NUT solution described above was
uplifted by A. Sen to ten dimensions (Nucl. Phys. {\bf 440} (1995) 421),
while C.V. Johnson and R.C. Myers have shown that the corresponding
extremal cases may be interpreted as exact string backgrounds using
the WZWN approach ({\it A Conformal Theory of a Rotating Dyon},
preprint PUPT--1524, McGill/95--01, hep-th/9503027). Further development
within the present framework can be found in
D.V. Gal'tsov, A.A. Garcia, and O.V. Kechkin,
{\it Symmetries of the Stationary Einstein--Maxwell Dilaton Theory},
hep-th/9504155, and D.V. Gal'tsov and O.V. Kechkin, {\it U--duality
and Symplectic Formulation of Dilaton-Axion Gravity}, hep-th/9507005.
}.

An integrability property of the EMDA theory with two commuting
Killing vectors opens the way for  application of a variety of generating
techniques to build (at least) as many classical solutions to this theory
as one has in General Relativity. Construction of EMDA classical
solutions seems to be useful in the search for the so-called
exact four-dimensional string backgrounds. We hope that the possibility
of getting (in principle) all classical solutions with two commuting
Killing
vectors will  also be helpful in understanding more general aspects of
this approach. \bigskip

\noindent {\bf Acknowledgments}

I wish to thank the Organizing Committee of the Heat Kernel Conference
and the Department of Physics, University of Manitoba, for hospitality
during this enjoyable meeting.
I am also grateful to the Physics
Department of the CINVESTAV del I.P.N., Mexico, for hospitality and to
CONACyT for support during my visit
while part of this work was done.
This work was supported in parts by the Russian Foundation
for Fundamental Research
grant 93--02--16977, and by the International Science Foundation
grant M79000.
\bigskip

\n {\bf References}
\medskip

\frenchspacing

\item{[1]}
R. Geroch, Journ. Math. Phys. {\bf 13}, 394 (1972);
 V.A. Belinskii and V.E. Zakharov. Sov. Phys. JETP, {\bf 48},
985 (1978); {\bf 50}, 1 (1979);
G. Neugebauer, {\it Journ. Phys. A: }{\bf 12}, L67; {\bf 1}, L19 (1979);
D. Maison, Journ. Math. Phys.
{\bf 21}, 871 (1979);
I. Hauser and F. J. Ernst, Phys. Rev. {\bf D 20}, 362, 1783
(1979).

\item{[2]}
D. Kramer, H. Stephani, M. MacCallum, and E. Herlt, {\it Exact
Solutions of the Einstein Field Equations\/}, CUP, 1980.

\item{[3]}
W. Kinnersley,  Journ. Math. Phys. {\bf 14}, 651 (1973);
{\bf 18}, 1529 (1977);
W. Kinnersley and D. Chitre, Journ. Math. Phys.
{\bf 18}, 1538 (1977); {\bf 19}, 1926, 2037 (1978);
G.A. Alekseev, Pis'ma Zh. Eksp. Teor. Fiz. {\bf 32}, 301 (1980);
P.O. Mazur, Acta Phys. Polon. {\bf B14}, 219 (1983);
A. Eris, M. G\"urses, and A. Karasu, Journ. Math. Phys.
{\bf 25}, 1489 (1984).

\item{[4]}
D. Maison, Gen. Rel. and Grav. {\bf 10}, 717 (1979);
V. Belinskii and R. Ruffini, Phys. Lett. {\bf B89}, 195 (1980).

\item{[5]}
B. Julia, in {\it Proceedings of the John
Hopkins Workshop on Particle Theory}, Baltimore (1981);
in {\it Superspace and Supergravity}, ed.\ by S. Hawking and M. Ro\v cek,
Cambridge, 1981;
in {\it  Unified Theories of More than Four Dimensions} ed.\ by
V. De Sabbata and E. Shmutzer, WS, Singapore 1983.

\item{[6]}
P. Breitenlohner and D. Maison, in {\it Solutions of the Einstein's
Equations: Techniques and Results}, ed.\ by C. Hoenselaers, W. Dietz,
Lecture Notes in Physics, {\bf 205} (1984) 276;
P. Breitenlohner, D. Maison, and G. Gibbons,
Comm. Math. Phys. {\bf 120}, 253 (1988).

\item{[7]}
H. Nicolai, Phys. Letts. {\bf B194} (1987) 402.

\item{[8]}
G. Neugebauer and D. Kramer, {\it Ann. der Physik (Leipzig)} {\bf
24}, 62 (1969); in {\it Galaxies, Axisymmetric Systems and
Relativity}, ed. by M. MacCallum, CUP, 1986, p.149.

\item{[9]}
G. Veneziano, Phys. Lett. {\bf B265} (1991) 287;
 K. Meissner and G. Veneziano, Phys. Lett. {\bf B267} (1991) 33;
M. Gasperini, J. Maharana and G. Veneziano, Phys. Lett. {\bf B272}
(1991) 277;
A. Sen, Phys. Lett. {\bf B271} (1991) 295; Phys. Lett. {\bf B274}
(1991) 34;
S.F. Hassan and A. Sen, Nucl. Phys. {\bf B375} (1992) 103;
A. Sen, Nucl. Phys. {\bf B404} (1993) 109.

\item{[10]}  A. Giveon, M. Porrati, and E. Rabinovici, Phys. Rept.
{\bf 244} (1994) 77.

\item{[11]}  A. Font, L. Iba\~nez, D. L\"ust, and F. Quevedo, {\it Phys.
Lett.} {\bf B249} (1990) 35;
S.J. Rey, {\it Phys. Rev.} {\bf D43} (1991) 526;
A. Sen, Nucl. Phys. {\bf B404} (1993) 109; Phys. Lett. {\bf 303B}
(1993) 22; Int. J. Mod. Phys. {\bf A8} (1993) 5079; 
Mod. Phys. Lett. {\bf A8} (1993) 2023;
J.H. Schwarz and A. Sen, Nucl. Phys. {\bf B411} (1994) 35; Phys.
Lett. {\bf 312B} (1993) 105.

\item{[12]}
D.V. Gal'tsov and O.V. Kechkin,  {\it Ehlers--Harrison Type Transformations
in Dilaton--Axion Gravity },  Phys. Rev. {\bf D50}
(1994) 7394.

\item{[13]}
W. Israel and G.A. Wilson, Journ. Math. Phys. {\bf 13} (1972) 865.

 \item{[14]}
G. Horowitz, {\it What is True Description of Charged Theory: Black Holes
and Black Strings}, UCSBTH-92-52, hep-th/9301008.

\item{[15]}
B.K. Harrison, Journ. Math. Phys. {\bf 9}, 1774 (1968).

\item{[16]}
G.W. Gibbons, Nucl.Phys. {\bf B204}, 337 (1982);
G.W. Gibbons and K. Maeda, Nucl. Phys. {\bf B298}, 741 (1988);
D. Garfinkle, G.T. Horowitz, and A. Strominger Phys. Rev. {\bf D43}, 3140
(1991); {\bf D45}, 3888 ({\bf E}) (1992).

\item{[17]}
F. Dowker, J. Gauntlett, D. Kastor, and J. Traschen,
Phys. Rev. {\bf D49}, 2909 (1994).

\item{[18]}
G. Neugebauer, Habilitationsschrift, Jena, FSU, 1969
(unpublished).

\item{[19]}
J. Ehlers, in {\it Les Theories Relativistes de la
Gravitation}, CNRS, Paris, 1959, p. 275.

\item{[20]}
R. Kallosh, D. Kastor, T. Ortin, and T. Torma,
{\it Supersymmetry and Stationary Solutions in Dilaton-Axion Gravity},
SU--ITP--94--12, hep-th/9406059.

\item{[21]}
A. Sen, Phys. Rev. Lett. {\bf 69}, 1006 (1992).

\item{[22]}
C.V. Johnson and R.C. Myers, {\it Taub--NUT Dyons in
Heterotic String Theory},
IASSNS--HEP--94/50, hep-th/9406069.

\item{[23]}
R. Kallosh and T. Ortin, Phys. Rev. {\bf D48}, 742 (1993).

\item{[24]}
C. Johnson, {\it Exact Models of Extremal Dyonic
Black Hole Solutions of Heterotic String Theory},
IASSNS-HEP-94/20, hep-th/9403192.

\item{[25]} D.V. Gal'tsov, {\it Integrable systems in stringy gravity},
Phys. Rev. Lett. {\bf 74} (1995) 2863.

\item{[26]}
I. Bakas, {\it O(2,2) Transformations and the String
Geroch Group}, CERN--TH--7144/94, hep--th/9402016.

\item{[27]} D.V. Gal'tsov and A.A. Garcia, in {\it Abstracts of Cornelius
Lanczos International Conference\/}, Dec. 1993, North Carolina, p.100.

\bye